\documentclass[12pt]{article}

\usepackage{color,amsmath,amssymb,exscale,epsfig}
\newcommand{\be}{\begin{equation}}
\newcommand{\bea}{\begin{eqnarray}}
\newcommand{\ee}{\end{equation}}
\newcommand{\eea}{\end{eqnarray}}

\usepackage{slashed,axodraw,epsfig} 
\usepackage{accents}

\title{
\Large
\textbf{
A charged Z' to conciliate the apparent disagreement between $t\bar t$ Tevatron forward-backward asymmetry and LHC charge asymmetry
}\vspace*{1.0cm}
}

\author{Ezequiel \'Alvarez$^{1,2,}$\footnote{sequi@df.uba.ar}\ \ and Estefania Coluccio Leskow$^{1,}$\footnote{ecoluccio@df.uba.ar}
\vspace{5mm}
\\
\normalsize\emph{(1) CONICET, IFIBA, Departamento de F\'{\i}sica, FCEyN, Universidad de Buenos Aires,} \\
\normalsize\emph{Ciudad Universitaria, Pab.~1, (1428), Ciudad de Buenos Aires, Argentina.} \\
\normalsize\emph{(2) CONICET, INFAP, Departamento de F\'{\i}sica, FCFMyN, Universidad Nacional de San Luis,} \\
\normalsize\emph{Av.~Ej\'ercito de los Andes 950, (5700), San Luis, Argentina.} \\
}

\date{}

\begin{document}
\setcounter{page}{0}
\maketitle

\vspace*{1cm}
\begin{abstract}
	We propose a charged, electrically neutral, and flavour changing $Z'$ model to conciliate the apparent disagreement between the important excess found in the $t\bar t$ Tevatron forward-backward asymmetry and the null --compatible with negative--  results found in the LHC charge asymmetry.  We show that this model contributes positively to the forward-backward asymmetry, whereas naturally a new cancellation is turned on at the LHC, yielding a null, or even negative, charge asymmetry.  We found the region in parameter space that is simultaneously allowed by the stringent Tevatron and LHC observables.  We show that the model is safe to atomic parity violation constraints and to $tj/\bar t j$ resonance searches, and propose a possible increase in the $Z'$ width to avoid $t\bar t j$ constraints.  We evaluate the constraints to the model, as well as distinctive features in the fore-coming experimental results.
\end{abstract}

\thispagestyle{empty}
\newpage

\setcounter{page}{1}

\section{Introduction}

The recent discovery of a new Higgs-like boson by the ATLAS~\cite{:2012gk} and CMS \cite{:2012gu} Collaborations at the LHC, with a mass of approximately $125$~GeV, is an exceptional step towards the verification of the Standard Model (SM). 
The SM has been tested by many experiments over the last decades and has successfully described high energy particle interactions. However, the electroweak symmetry breaking mechanism has not been yet understood. The LHC will be extensively devoted to this subject and to measuring the properties of the new particle in order to explore the underlying theory from which it arises. The understanding of this new particle interactions could be an important probe of New Physics (NP) in coming years. 

Another particle sensitive to NP is the top quark \cite{topNP}, not only because its mass is close to the electroweak symmetry breaking scale, but also because of its relatively little exploration. Experimental results that could give us hints of NP effects in this sector have been reported \cite{cdfviejo,tevatron}, being the $p\bar p \to t\bar t$ forward-backward asymmetry ($A_{FB}$) measurement, probably one of the most remarkable ones. Both CDF and D0 Collaborations measured the $t \bar t$ cross section ($\sigma_{t\bar t}$) in good agreement with the SM \cite{1201.6653}, however there exists discrepancy in the $A_{FB}$ between the theory and the experimental results. This asymmetry enables the study of the top pair production mechanism and it is customary to define it as 
\bea
A_{FB}=\frac{N(\Delta y>0)-N(\Delta y<0)}{N(\Delta y>0)+N(\Delta y<0)},\label{eq:afb}
\eea
where $\Delta y=y_{t}-y_{\bar{t}}$ is the difference in rapidity of top and anti-top quarks along the proton momentum direction. 

While the SM prediction for $A_{FB}$ at NLO in QCD is $0.087 \pm 0.01$ \cite{AfbTeva2}, 
results from CDF and D0 report excesses in their measured asymmetries already from the first published results \cite{cdf1} in 2008. The most recent CDF results give an inclusive parton level asymmetry $A_{FB}=0.162 \pm 0.047$ \cite{AfbCDF} in agreement with an independent D0 measurement of $A_{FB}=0.196 \pm 0.065$ \cite{AfbD0}. The largest disagreement with the SM $A_{FB}$ prediction was announced this year by CDF in the differential measurements for $A_{FB}(M_{t\bar t})$ and $A_{FB}(|y_t-y_{\bar t}|)$ \cite{AfbCDF}.  The fitted results of these differential measurements have a p-value statistical significance of $p=0.006$ and $p=0.008$, respectively.

Many NP models \cite{ttbar} arose to account for the excess measured in the $A_{FB}$. If this excess is generated by new physics, then these models could be tested at the LHC. Since this machine is a symmetric $p p$ collider, the top quark forward-backward asymmetry vanishes. However, an asymmetry in charge ($A_{C}$) can be measured and it is defined by
\bea
A_{C}=\frac{N(\Delta |y|>0)-N(\Delta |y| <0)}{N(\Delta |y|>0)+N(\Delta |y| <0)}.\label{eq:ac}
\eea

The current experimental values for $A_C$ are $A_{C}=0.029\pm 0.018\pm 0.014$ at ATLAS~\cite{atlasacy} and $A_{C}=0.004\pm 0.010\pm 0.011$ at CMS \cite{cmsacy}, both consistent with the SM prediction of $0.0115 \pm 0.0006$~\cite{smafbac}. Almost all the models that tried to explain the large $A_{FB}$ also predicted a large value for $A_{C}$ and as a result most of them were excluded.

According to the nature of the new particle exchange, these models fall mainly into two sets: those with new {\it s-}channel processes and those with a new {\it t-}channel exchange mediator. Many of these models have already been discarded not only due to $A_{FB}$ and $A_{C}$, but also to other precision LHC measurements. For instance, dijet observables \cite{1108.6311,1103.3864} have excluded many {\it s-}channel models, while {\it t-}channel ones such as flavour changing neutral current (FCNC) $Z'$ models \cite{Jung:2009jz,1003.3461,Barger:2011ih,1104.0083} have been discarded by same-sign top pair production \cite{Berger:2011ua,Berger:2011sv}. In order to avoid this last constraint, models with a charged $Z'$ and/or a $W'$ \footnote {Along this work $Z'$ refers to an {\it electrically neutral} boson and $W'$ to an electrically charged one.} arose \cite{1207.0643,1208.4675,1203.1320}.  An example of this kind of models is an specific one \cite{jung} where a horizontal gauge symmetry yields a flavour-changing and a flavour-conserving neutral boson which has been discarded by atomic parity violation (APV) observables \cite{1203.1320}. 

In this work \cite{ichep} we study a phenomenological charged $Z'$ model with flavour violating couplings to $u$ and $t$ quarks.  We stress that the new boson is electrically neutral. This $Z'$ has a mass larger than the top mass and no other partner coming from gauge invariance \cite{1203.1320,jung}. The reasons for this phenomenological model come out to be two-folded: {\it (i)} constraints as FCNC top decays and same-sign top production are avoided, whereas APV constraints are largely relaxed; and {\it (ii)} it appears a cancellation in $A_{C}$ which is not present in $A_{FB}$, yielding a possible explanation for the apparent disagreement between these observables.

This model could solve the apparent disagreement between $A_{FB}$ and $A_{C}$ in an innovative way. In most of the models that try to account for the large $A_{FB}$ measured at the Tevatron, the excess in this asymmetry also implies an excess in $A_{C}$, and the agreement is sought as an intermediate balance in which $A_{C}$ is not too large while $A_{FB}$ is not too small. In the model presented in this work, on the other hand, the agreement in some part of the parameter space has to be sought as making $A_{FB}$ large without making $A_{C}$ too negative.

We study the Tevatron and LHC phenomenology of this model and verify that the cancellation takes place, making possible the simultaneous explanation not only of both $A_{FB}$ and $A_{C}$, but also of all CDF unfolded results, APV, and LHC $t \bar t$ cross section within the $95~\%$ C.L. However, the model predicts an excess in $t \bar t j$  final state. To avoid this difficulty, we explore the possibility of increasing the $Z'$ width, assuming that the $Z'$ decays to not detectable particles a fraction of the times. Although it is not the purpose of this work to address the fate of the invisible decay, we mention that these particles could be, for instance, dark matter or neutrinos.  This new feature of the model predicts single top production with a particular topology, which we also explore.

This work is divided as follows.  In the next section we present the model and its phenomenology and explain how the cancellation in $A_C$ takes place at the LHC.  In section 3 we perform Monte Carlo simulations of Tevatron and LHC and we find the region in parameter space compatible with all the constraints.  In section 4 we discuss constraints and predictions for the model, and section 5 contains the conclusions.

\section{Phenomenology of a charged $Z'$ model}

In this section we present the Lagrangian of a phenomenological $Z'$ model together with a description of its contributions to the $t\bar t$ forward-backward asymmetry at the Tevatron and charge asymmetry at the LHC. We find the expected constraints to the model, which are analysed in section 4.

\subsection{The Model}
We consider a model containing a charged, spin-one, colorless particle with flavour-violating interactions which we call $Z'$. We assume this particle couples only to right-handed {\it u} and {\it t} quarks since the left-handed coupling is constrained by {\it B-} physics \cite{Zhu:2011ww,Duraisamy:2011pt}. We also assume that its mass is larger than the top mass, avoiding a flavour-changing top decay \cite{cmspastop028}. The phenomenological NP Lagrangian is then given by:

\bea
{\cal L}_{NP}=f_{R}\bar u \gamma^{\mu} P_{R} t Z'_{\mu}+ f_{R} \bar t \gamma^{\mu} P_{R} u Z'^{\dagger}_{\mu},
\eea
where $P_{R}=\frac{(1+\gamma^{5})}{2}$ and $f_{R}$ is the right-handed coupling.

It is important to note that we are considering a charged boson, so that $Z'_{\mu}$ is not the same particle as its conjugate partner, $Z'^{\dagger}_{\mu}$. Under this condition, the production of same sign tops is forbidden. Models in which these two particles are the same particle, i.e models with {\it neutral} $Z'$ bosons, allow the production of same sign top pairs and as a consequence, are excluded \cite{zneutra}. 

\subsection{Phenomenology for Tevatron and LHC $t \bar t$ asymmetries} \label{pheno}

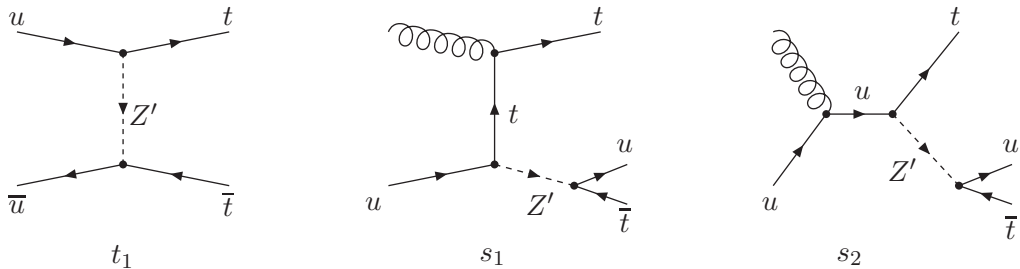
\begin{figure}[!ht]
\begin{center}\begin{picture}(410,100)(0,-60)
\small
\ArrowLine(10,10)(50,2) \Text (10,15)[]{$u$}
\Vertex(50,2){1.5} 
\ArrowLine(50,2)(90,10) \Text (90,16)[]{$t$}
\DashArrowLine(50,2)(50,-40){2} \Text (58,-21)[]{$Z'$}
\ArrowLine(50,-40)(10,-48) \Text (10,-55)[]{$\overline u$}
\Vertex(50,-40){1.5} 
\ArrowLine(90,-48)(50,-40) \Text (90,-55)[]{$\overline t$}
\Text(50,-75)[]{{${t_{1}}$}}

\Gluon(150,10)(190,2){4}{5} 
\Vertex(190,2){1.5} 
\ArrowLine(190,2)(230,10) \Text (230,16)[]{$t $}
\ArrowLine(190,-40)(190,2) \Text (198,-21)[]{$t$}
\ArrowLine(150,-48)(190,-40) \Text (145,-55)[]{$u$}
\Vertex(190,-40){1.5} 
\DashArrowLine(190,-40)(220,-48){2} \Text (208,-55)[]{$Z'$}
\Vertex(220,-48){1.5} 
\ArrowLine(220,-48)(240,-40) \Text (240,-33)[]{$u$}
\ArrowLine(240,-55)(220,-48) \Text (240,-61)[]{$\overline t $}
\Text(190,-75)[]{{$s_{1}$}}

\Gluon(295,10)(315,-21){4}{5} 
\Vertex(315,-21){1.5} 
\ArrowLine(295,-48)(315,-21) \Text (295,-55)[]{$u$}
\ArrowLine(315,-21)(340,-21) \Text (330,-13)[]{$u$}
\Vertex(340,-21){1.5}
\ArrowLine(340,-21)(365,10) \Text (365,16)[]{$t$}
\DashArrowLine(340,-21)(365,-48){2} \Text (345,-42)[]{$Z'$}
\Vertex(365,-48){1.5} 
\ArrowLine(365,-48)(385,-40) \Text (386,-33)[]{$u$}
\ArrowLine(385,-55)(365,-48) \Text (386,-64)[]{$\overline t$}
\Text(325,-75)[]{{$s_{2}$}}
\end{picture}
\vspace{.5 cm}
\caption{\footnotesize Feynman diagrams for $pp,p \overline p \to t \overline t (u)$ involving a $Z'$: In $t_{1}$ the $Z'$ is exchanged through a {\it t-}channel and in $s_{1}$ and $s_{2}$ the $Z'$ goes through an {\it s-}channel. We show that $s_1$ cancels the contribution to the charge asymmetry of $t_1$ at the LHC.} \label{feyn}
 \end{center}
\end{figure}

The Feynman diagrams for $pp,p \bar p \to t \bar t (u)$ involving a $Z'$ boson in the model described previously are shown in Fig.~\ref{feyn}. We denote by $t_{1}$ the diagram where this particle is exchanged through a {\it t-}channel and by $s_{1}$ and $s_{2}$ those diagrams where the $Z'$ goes through an {\it s-}channel. In the former case, the $Z'$ contributes to a $t\bar t$ final state, while in the later, to $t\bar t u$ production. Since $Z'\neq Z'^{\dagger}$, $s_{1}$ and $s_{2}$ have different conjugate diagrams, $\bar s_{1}$ and $\bar s_{2}$, which at the Tevatron, due to the symmetry in $p \leftrightarrow \bar p$, have the same strength as $s_{1}$ and $s_{2}$. On the contrary, at the LHC $\sigma(\bar s_{1}, \bar s_{2}) \ll \sigma(s_{1},s_{2})$. 

The cornerstone of our analysis is the observation that at the LHC there is a cancellation of the charge asymmetry coming from the contributions of the {\it t-} and {\it s-}channel processes, explaining the small and compatible with negative charge asymmetry measured by this experiment. This cancellation is not present at the Tevatron where as a matter of fact a large $A_{FB}$ has been measured. We see in the following paragraphs how the {\it t-}channel diagram contributes positively to the asymmetries while at the LHC the {\it s-}channel ones have a negative contribution.

To understand this cancellation it is important to clarify two points. First, the reason why the {\it t-}channel contributes positively to both the $A_{C}$ and the $A_{FB}$ asymmetries whereas the {\it s-}channel contribution is negative and only noticeable at the LHC. Second, why the {\it t-}channel process is privileged at the Tevatron while the {\it s-}channel is turned on at the LHC. 

To study the first point one should first realize that the $s_{2}$ process is suppressed with respect to $s_{1}$ and $t_{1}$ since the up quark propagator carries all the energy of the process. We can then compare $t_{1}$ and $s_{1}$ (which is a {\it t-}channel diagram if thought up to $t Z'$ final state) by studying the general dynamics of a {\it t-}channel process. 

In a general {\it t-}channel $1,2 \to 1',2'$ process, where the same number indicates a shared vertex, the relevant factor coming from the propagator of an exchanged $X$ particle is,

\bea
\frac{1}{(p_{1}-p_{1'})^{2}-m^{2}_{X}}&=&\frac{1}{m^{2}_{1}+m^{2}_{1'}-2E_{1}E_{1'}+2\vec p_{1}.\vec p_{1'}-m^{2}_{X}}.
\label{dynamics}
\eea
In general $m_{1}$ can be neglected. For the case $m_{1'}<m_{X}$, ($t_{1}$ in Fig.$~\ref{feyn}$), the events with the largest cross sections are those where $\vec p_{1}.\vec p_{1'}>0$. For the sake of brevity we refer to this condition as 1 and 1' having the same direction and to $\vec p_{1}.\vec p_{1'}<0$ as having opposite direction. If $m_{1'}>m_{X}$ ($s_{1}$ in Fig.$~\ref{feyn}$) the same holds unless $E_{1}$ is too small, although it can be seen in the center of mass system that this is kinetically forbidden for this specific process. Note that although there will also be contributions coming from the Lorentz structure of the vertices, the only analysis of the dynamic in Eq.~(\ref{dynamics}) already results in a good approach to compare diagrams $t_{1}$ and $s_{1}$.


From this reasoning we see that in the {\it t-}channel diagram $t_{1}$ of Fig.$~$\ref{feyn}, the top quark is likely to have the same direction as the incoming up quark, contributing to a {\it positive} asymmetry. 

Following the same logic, in the {\it s-}channel diagram $s_{1}$ of Fig.$~$\ref{feyn}, the $Z'$ boson is the one that tends to have the same direction as the incoming up quark and transmits it to its decay products $\bar t$ and $u$. This results in a {\it negative} contribution to the asymmetry. At this point it is interesting to note that at the Tevatron $s_{1}$ and its conjugate contribute the same, however at the LHC $s_{1}$ dominates over $\bar s_{1}$ and as a result the net contribution from these two diagrams to the charge asymmetry is negative and that is why the {\it s-}channel effectively contributes to the asymmetry only in this experiment.


The second point to analyse involves two questions: why the {\it s-}channel is turned on at the LHC and why the {\it t-}channel process dominates at the Tevatron. The first one has to do with the energy of the accelerator: since at the LHC the phase space is larger than at the Tevatron, the {\it s-}channel, which has a $Z'$ on-shell, is turned on in this machine resulting in a cancellation of the charge asymmetry when all the processes are considered. The second one concerns the nature of the collisions at the Tevatron: the {\it t-}channel process is privileged because it involves antiquarks, present in the colliding antiprotons, so its positive contribution is enhanced in the forward backward asymmetry measured in this accelerator. 

Summarizing, let us remark once again that the {\it s-}channel, having a negative contribution to the asymmetry, is crucial for the cancellation of the charge asymmetry and thus for the simultaneous explanation of the forward-backward and charge asymmetry measurements.


\subsection{Expected constraints to the model}\label{constraints}

We refer in this subsection to the expected constraints to the model in a qualitative way. We study all of them in some depth in section \ref{results} and \ref{discussion}. 

A direct constraint to the model comes from $tj/\bar t j$ resonance searches. Apart from our model, many other models of NP \cite{1102.0018,1106.3086,1111.5857} predict a resonance in the $tj/\bar t j$ system of $t \bar tj$ final state. We analyse in section \ref{discussion} the experimental results in order to set limits to our model coming from these resonances.

One of the indirect constraints to the model comes from the limits in $t\bar t j$ production, since the {\it s-}channel processes contribute to it. In order to relax this limit, we propose new decays for the $Z'$ such as dark matter or neutrinos. These new decays imply an increment of the $Z'$ width which only affects the {\it s-}channel processes; the $t_{1}$ process is not altered since the $f_{R}$ coupling remains the same and $Z'$ is not on shell in this channel.

Observe that the increment of the $Z'$ width caused by the new invisible decays of this particle, results in a particular single top production topology. In fact, when the $Z'$ decays to undetectable particles, the final state will be a top, missing energy and no $b$-jets. 

Another indirect constraint comes from APV. The $Z'tu$ vertex generates one loop corrections to the $Zuu$ effective coupling that affect low-energy precision tests of parity-violating observables \cite{1203.1320,Dzuba:2012kx}. The strongest constraints come from APV measurements in cesium~\cite{Wood:1997zq}. We investigate the parity-violating atomic transitions sensitive to the nuclear weak charge within this model and show the results in section \ref{results} and \ref{discussion}.

\section{Numerical Results} \label{results}
The analysis of the previous section led us to the understanding of the charge asymmetry cancellation mechanism, which makes possible the simultaneous explanation of the forward-backward and charge asymmetry experimental results at the Tevatron and the LHC, respectively. In this section we search numerically for this cancellation and investigate the allowed parameter space of the model by confronting it with many relevant observables and with the major constraints discussed in the previous section. The parameter space considered is delimited by $200$ GeV $<M_{Z'}<500 ~ \rm{GeV}$ and $0.5<f_{R}<1.2$.

Using {\sc MadGraph5} \cite{MG} we simulate $t \bar t (u)$ production at the Tevatron and the LHC@7$\rm TeV$ within the $Z'$ model at parton level according to the diagrams of Fig.$~\ref{feyn}$ and their conjugates, in addition to the SM LO $ t \bar t$ contribution. Since in the SM these processes do not generate a charge asymmetry, the $A_{C}$ computed with the simulated collisions contains NP contributions only. Hence, in order to compute the model predictions, it is necessary to include the SM@NLO contribution to $A_{C}$.

If the NP contribution to the total cross section is small, $\sigma_{SM} \gg \sigma_{NP}$ (where $\sigma_{NP}$ contains both SM-NP interference and NP squared contributions), we can approximate the asymmetry by \cite{Alvarez:2010js}

\be
A_{C} \approx A_{C}^{NP+SM@LO}+A_{C}^{SM@NLO}.
\ee

We study the $Z'$ model in the parameter space previously mentioned confronting it with the last differential measurements of $A_{FB}$ \cite{AfbCDF} and $\sigma_{t \bar t}$ \cite{cscdf} at parton level from CDF, and $\sigma_{t \bar t}$ \cite{CMS2} and $A_{C}$ \cite{cmsacy} from CMS. We use CDF results since their discrepancy with the SM has larger statistical significance than those of D0 \cite{AfbD0}, and the main purpose of this work is to present a model capable of reconciling two measurements ($A_{FB}$ and $A_{C}$) which may seem to be in disagreement. On the other hand, we use CMS measurements because they yield the most precise results.
We perform a $\chi^{2}$ test with all the observables measured at CDF and confront the model with all the other ones in a separate way each. We analyse each of these constraints in the following paragraphs.

The last measurement of $A_{FB}$ published by CDF \cite{AfbCDF} shows $A_{FB}$ as a function of both the invariant mass $M_{t \bar t}$ and $\Delta y$. The ranges for the bins used in that analysis, and in our $\chi^{2}$ test, are: $[0-450;450-550;550-650;650-\infty] ~ \rm{GeV}$ for $M_{t \bar t}$, and $[0-0.5;0.5-1;1-1.5;1.5-\infty]$ for $\Delta y$. By requiring the p-value to be greater than 0.05 we select the points in parameter space which are in agreement with CDF results at a 95$\%$ C.L. The $M_{Z'}$ vs. $f_{R}$ region consistent with Tevatron measurements is delimited by the green dashed lines present in all the figures that follow in this section. 

To confront our model against the measurement of the inclusive $t\bar t$ cross-section at LHC, we do as follows.  We take as the experimental input for the inclusive LHC@7TeV $t \bar t$ cross section, the CMS combination which is $165.8 \pm 13.3 ~\rm pb$ \cite{CMS2}; whereas for the theoretical input we use the calculation made with HATHOR, which gives $164 ^{+11}_{-16} ~ \rm pb$ \cite{hathor}.  Since their central values agree, and their errors summed in quadrature represent a $13$\% of the cross-section, we test the cross-section in the simulations of our model against a similar simulation with only SM and we set the error to be the $13$\% of the cross-section.    It can be shown that this procedure is equivalent to using a K-factor. Since our NP final state goes up to $t\bar t j$, we do the SM corresponding simulation and use Pythia to account for initial and final state radiation and the MLM \cite{mlm} matching scheme to avoid double counting.  

We first analyse, separately, the positive and negative contributions to the charge asymmetry discussed in the previous section with the only purpose of explicitly observing each of them. 

We define the {\it t-} and {\it s-}channel charge asymmetries, $A_{C_{t}}$ and $A_{C_{s}}$,

\bea
A_{C_{t}}=\frac{N^{+}(t,SM)-N^{-}(t,SM)}{N^{+}(t,SM)+N^{-}(t,SM)+N^{+}(s)+N^{-}(s)},
\eea

\bea
A_{C_{s}}=\frac{N^{+}(s)-N^{-}(s)}{N^{+}(t,SM)+N^{-}(t,SM)+N^{+}(s)+N^{-}(s)},
\eea
where $N^{+(-)}(t,SM)$ is the number of events with a positive (negative) value of $\Delta|y|$ when the {\it t-}channel and the SM processes at tree level are considered, while $N^{+(-)}(s)$ denotes the same quantity except that in this case only the {\it s-}channel processes are taken into account. With these definitions the charge asymmetry of NP+SM@LO is given by

\bea
A_{C}^{NP+SM@LO}=A_{C_{t}}+A_{C_{s}}. \label{suma}
\eea

\begin{figure}[!h]
\begin{center}
\includegraphics[width=0.8 \textwidth]{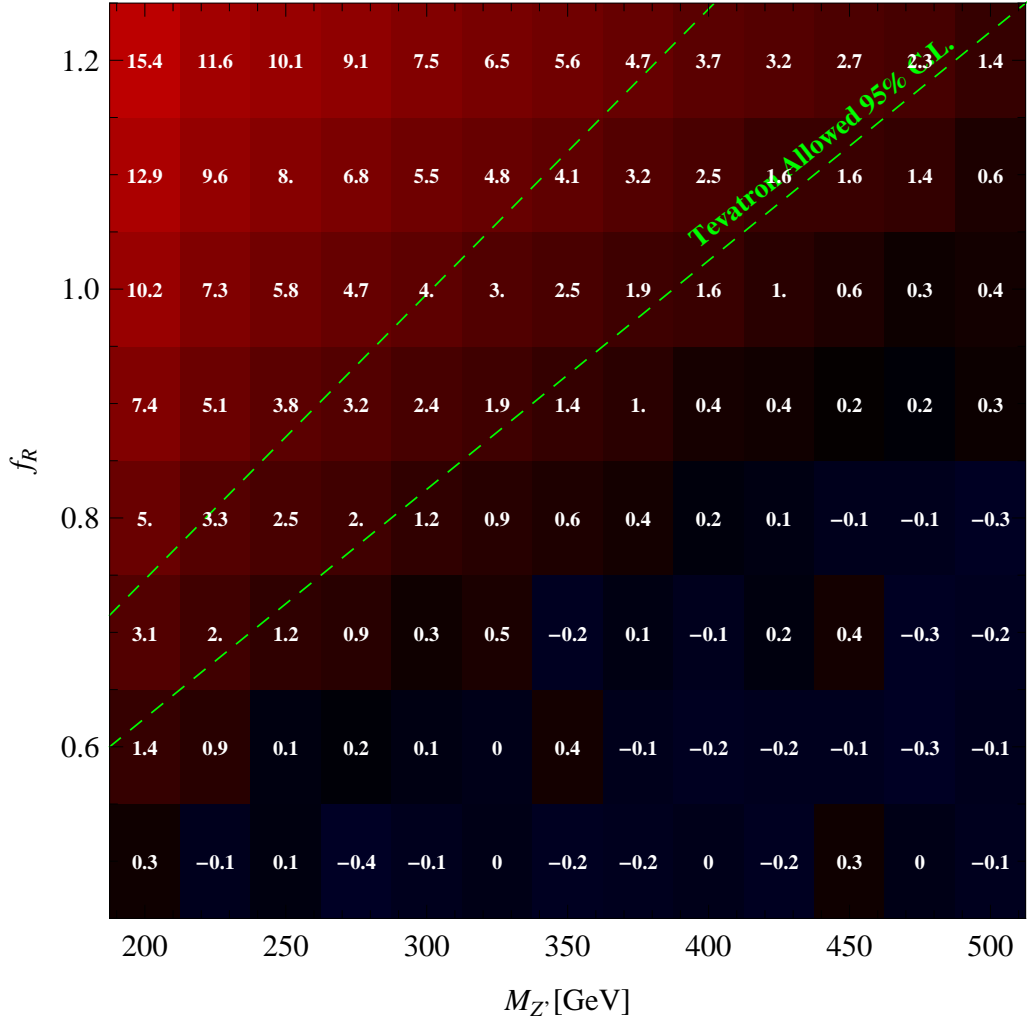}
\caption{\footnotesize [Color online] {\it t-}channel contribution to the charge asymmetry. The background colours indicate the sign of the contribution for every point in the parameter space; red (blue) represents positive (negative) sign. The tone of the colours stands for the absolute value of the contribution; the more intense the tone, the larger the absolute value. The numbers in every point are the difference of $A_{C_{t}}+A^{SM@NLO}/2$ to half the measured value of the charge asymmetry, in units of the experimental error. The green dashed lines define the region consistent with Tevatron limits at 95\% C.L.}\label{tchannel1}
\end{center}
\end{figure}
\begin{figure}[!h]
\begin{center}
\includegraphics[width=0.8 \textwidth]{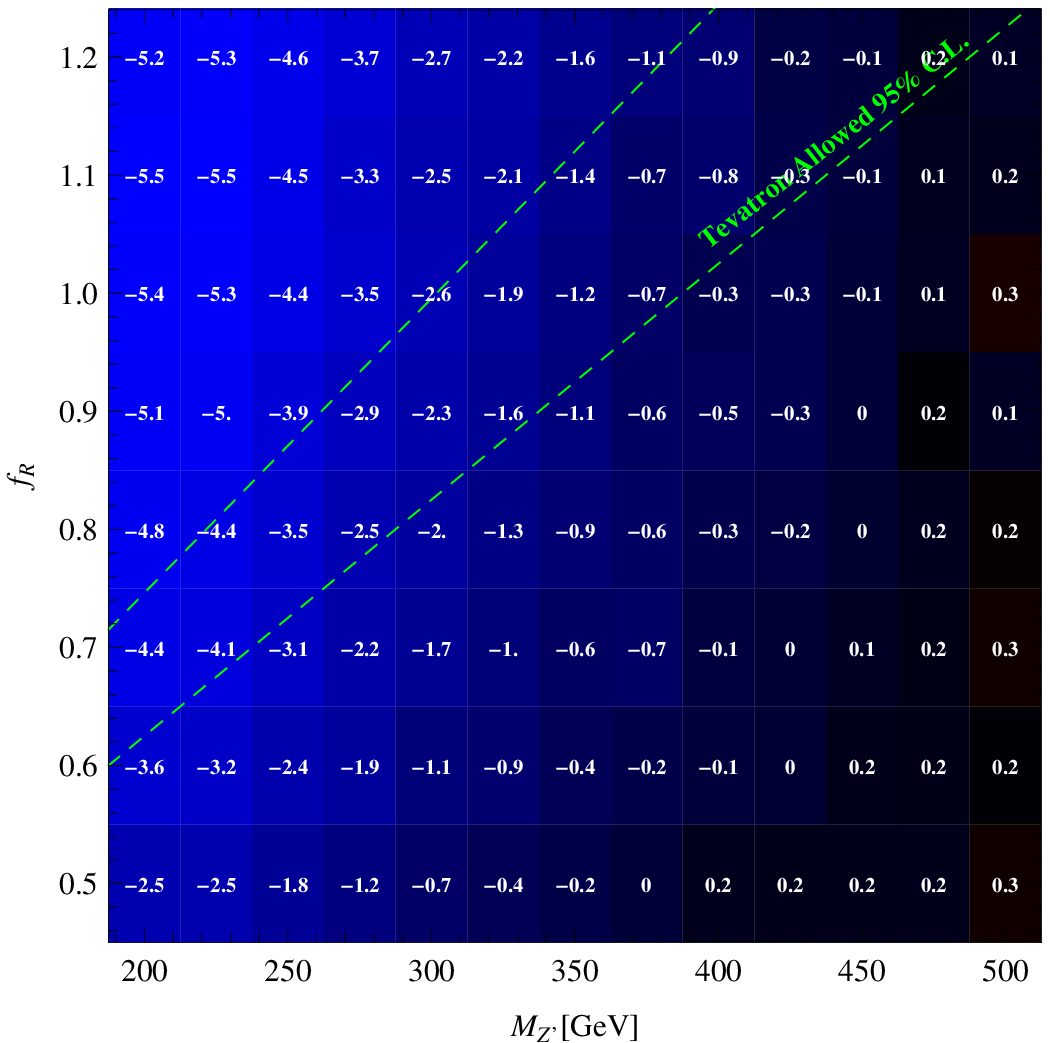}
\caption{\footnotesize [Color online] The same as the previous figure but for the {\it s-}channel contribution to the charge asymmetry.}\label{schannel1}
\end{center}
\end{figure}

We show in Fig.$~\ref{tchannel1}$ and Fig.$~\ref{schannel1}$, the {\it t-} and {\it s-}channel contributions to the charge asymmetry respectively. The background colours in the plots indicate the sign of the contribution for every point in the parameter space; red (blue) represents positive (negative) sign. The tone of the colours stands for the absolute value of the contribution; the more intense the tone, the larger the absolute value. The numbers in every point are the difference of $A_{C_{t}}+A^{SM@NLO}/2$ (Fig.$~\ref{tchannel1}$) and $A_{C_{s}}+A^{SM@NLO}/2$ (Fig.$~\ref{schannel1}$) to half the measured value of the charge asymmetry, in units of the experimental error. Note that these plots clearly exhibit the cancellation of the charge asymmetry in the region defined by Tevatron limits. In Fig.$~\ref{tchannel1}$ the differences to the measured value are mainly positive while those in Fig.$~\ref{schannel1}$ are mainly negative, what results in the expected cancellation of the charge asymmetry discussed in the previous section.

\begin{figure}[!h]
\begin{center}
\includegraphics[width=0.8 \textwidth]{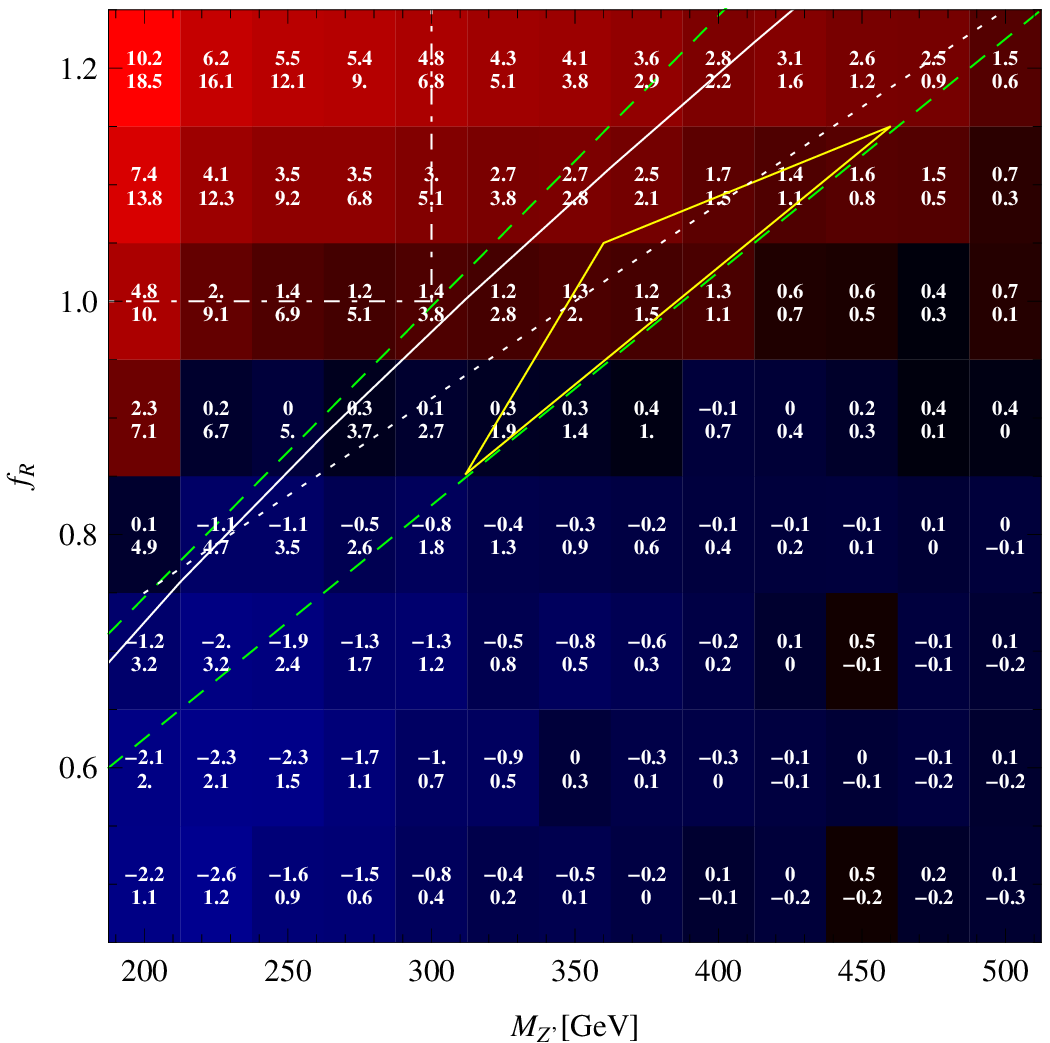}
\caption{\footnotesize [Color online]  {\it t-} and {\it s-}channel contributions to the charge asymmetry. In each cell, the upper number is the difference of $A_{C_{t}}+A_{C_{s}}+A_{C}^{SM@NLO}$ to the measured value of the charge asymmetry. The number below is the difference of $\sigma_{t \bar t}^{SM+NP}$ to the inclusive measured value of $\sigma_{t \bar t}$ at the LHC, as discussed in the text. The area delimited by the triangle contains the points consistent with Tevatron limits in which these two observables differ in less than 2 from their corresponding experimental values in units of the experimental error. Tevatron limits are defined by the dashed lines; APV excludes the region above the thick line. The dot-dashed lines limit the region excluded by $tj/\bar t j$ resonance searches by CDF while the region above the dotted line corresponds to the same searches by ATLAS. These constraints are discussed in section \ref{discussion}.}\label{all1}
\end{center}
\end{figure}

We show in Fig.$~\ref{all1}$ the contributions of both $A_{C_{t}}$ and $A_{C_{s}}$ using the same convention of colours and tones as in the two previous figures with the distinction that now in every cell there are two numbers. The upper one is the difference of $A_{C_{t}}+A_{C_{s}}+A_{C}^{SM@NLO}$ to the measured value of the charge asymmetry, in units of the experimental error. The number below is the difference of the model prediction for the $t\bar t$ inclusive cross-section to the measured value, in units of the error, as previously explained.
The area delimited by the triangle contains the points consistent with Tevatron limits in which these two observables differ in less than 2 from their corresponding experimental values in units of the experimental error. The dot-dashed lines limit the region excluded by $tj/\bar t j$ resonance searches by CDF while the region above the dotted line corresponds to the same searches by ATLAS. The parameter space above the thick line shows the APV excluded region. We discuss these constraints in the next section.

\begin{figure}[!h]
\begin{center}
\includegraphics[width=0.8 \textwidth]{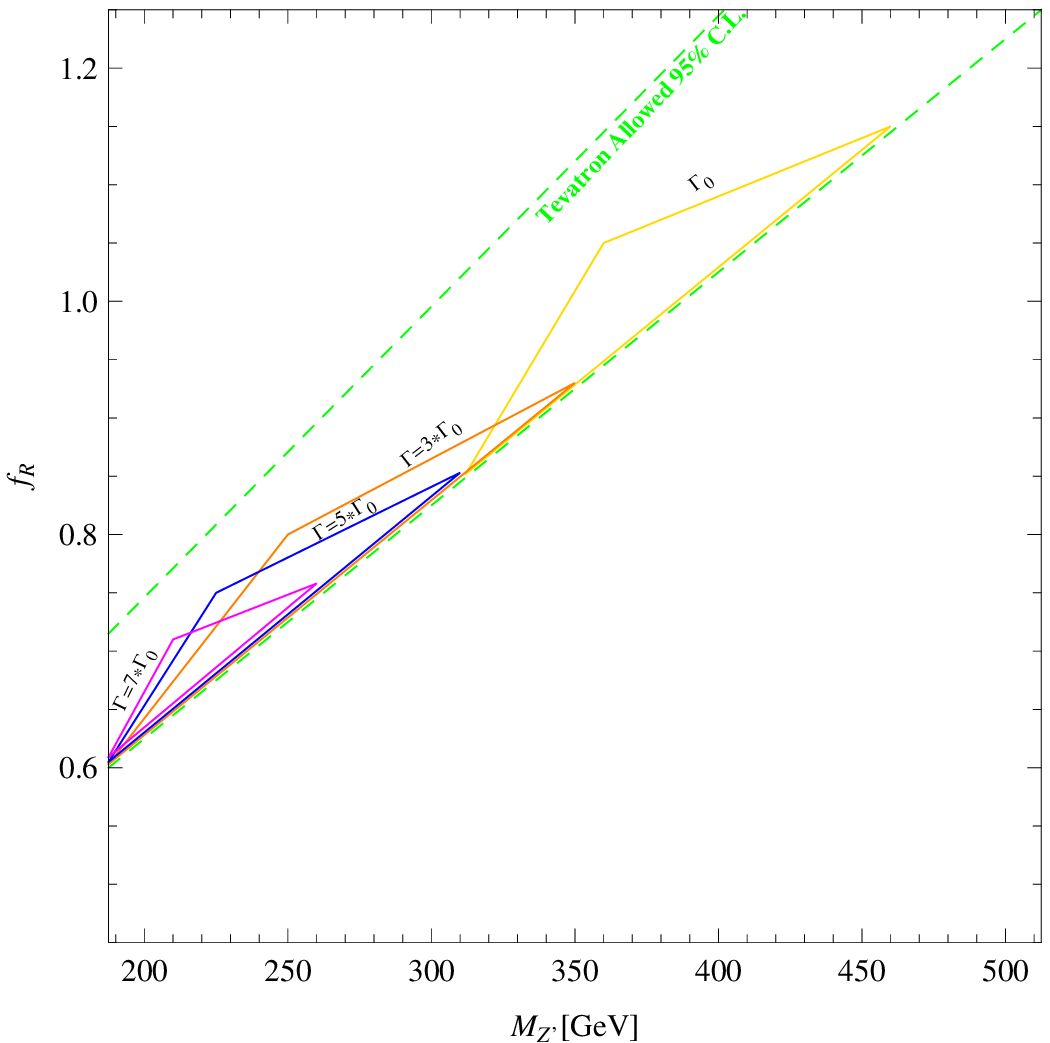}
\caption{\footnotesize  [Color online] Regions of the parameter space compatible with all the observables considered in this work from the Tevatron and the LHC for a $Z'$ with a width three (orange triangle), five (blue triangle) and seven (magenta triangle) times its value when the decay is solely to $u$ and $\bar t$ ($\Gamma_{0}$). We also show the yellow triangle of Fig.$~\ref{all1}$ that corresponds to no change in the $Z'$ width. This increment in the width avoids $t\bar t j$ constraints on the model.}\label{357}
\end{center}
\end{figure}

We have mentioned in section $\ref{constraints}$ that one indirect constraint to the model comes from $t\bar tj$ production. Since there are not available works on $t\bar tj$ limits that could be adapted to our model, we use in next section $W't d$ production results from Ref. \cite{1203.4489} as a rough estimation of the $t\bar tj$ production at the LHC in our model. We found that the width of the $Z'$ should be increased to avoid $t\bar t j$ constraints. We repeated the simulations for values of the $Z'$ width increased by three different factors and searched again for the allowed parameter space in these cases. We show in Fig.$~\ref {357}$ the allowed region for a $Z'$ width three (orange), five (blue) and seven (magenta) times its value when the decay is solely to $u$ and $\bar t$, which we denote by $\Gamma_{0}$. We also show in this plot the yellow triangle of Fig.$~\ref{all1}$ that corresponds to no change in the $Z'$ width. We checked that the narrow width approximation holds for all the values of the $Z'$ width considered.

The first remark concerning Fig.$~\ref{357}$ is that, as it can be seen, Tevatron results are not affected by the $Z'$ width modification. 
The second one has to do with the width increment effect on the allowed parameter space. Let us now analyse this point and start by investigating why the allowed region defined by the yellow triangle in Fig.$~\ref{all1}$, (also shown in Fig.$~\ref{357}$), gets excluded when larger values of the $Z'$ width are considered.

When the $Z'$ width is increased, the allowed areas of the parameter space appear displaced downward in Fig.$~\ref{357}$, to smaller values of $M_{Z'}$ and $f_{R}$, relative to the yellow triangle. This can be understood by looking at Fig.$~\ref{all1}$. The numbers in the cells inside the triangle that correspond to the larger values of $M_{Z'}$ and $f_{R}$ are those where the difference of $A_{C}$ to the measured value in units of the experimental error is closer to 2 compared to any other point in the triangle. These points are thus sensitive to getting excluded by any change in the model that could cause an increment in $A_{C}$. In fact, this is the case: when the $Z'$ width becomes larger, the proportion of processes in the {\it s-}channel decreases and, therefore, the negative contribution from $A_{C_{s}}$ to $A_{C}$ becomes smaller in absolute value. This translates into an increment of $A_{C}$ that causes a deviation from the measured value in more than 2 units of the experimental error in the upper region of the triangle. As a result, those points get excluded when the $Z'$ width is increased.

With a similar argument, but this time concerning $\sigma_{t \bar t}$, it can be explained why parts of the excluded region in Fig.$~\ref{all1}$ become allowed in Fig.$~\ref{357}$. In this case the sensitive observable is $\sigma_{t \bar t}$, which decreases as the $Z'$ width increases. The points inside the orange, blue and magenta triangles in Fig.$~\ref{357}$ are those where the difference of $\sigma_{t \bar t}$ to the measured value in units of the experimental error are larger than 2 in Fig.$~\ref{all1}$ and that is why they are excluded in this figure. However, they become allowed when the width is increased since this makes $\sigma_{t \bar t}$ decrease.

Conclusively, in a given allowed region, either in Fig.~\ref{all1} or \ref{357}, $A_C$ is the most sensitive observable in the sector of large $M_{Z'}$ and $f_R$, and $\sigma_{t\bar t}$ in the sector of smaller $M_{Z'}$ and $f_R$.


Finally, we see that the triangles in Fig.$~\ref{357}$ become smaller with larger values of the $Z'$ width suggesting that it cannot be increased arbitrarily because the effect of this increment on either $A_{C}$ or $\sigma_{t \bar t}$ (or both) eventually becomes important enough so as to exclude most of the parameter space.

In the next section we discuss the major constraints to the model, as well as its distinctive features.

\section{Constraints and predictions of the model} \label{discussion}

We have studied a model that simultaneously explains the apparently incompatible values of $A_{FB}$ and $A_C$. In this section we discuss the constraints to the model, its possible issues, and its distinctive features in the fore-coming experimental results.

\subsection{Direct constraint to the model}

We study in the following paragraphs the direct constraint to the model coming from $tj/\bar t j$ resonance searches in $t \bar t j$ final states. Tevatron and LHC experiments have looked for this resonance as a possible explanation for the forward-backward asymmetry measurements at the Tevatron. 

The first direct search for a particle $X$ that would give a $tj/\bar t j$ resonance in $t \bar t j$ events was made by CDF in Ref. \cite{1203.3894}. In this work they set upper limits at $95 \%$ C.L on $t \bar tj$ production via the new resonance particle $X$, as a function of the resonance mass for couplings $g_L=0$ and $g_{R}=1$. Fig. \ref{all1} shows the region of the parameter space excluded by this CDF search.

The CMS Collaboration also recently performed a search for a $W'$ boson via the process $dg\to tW', W' \to \bar t d$ \cite{cmstj}. The data showed no significant deviation from the standard model prediction and the $W'$ model with $g_{L}=0$ and $g_{R}=2$ was excluded for a $W'$ mass below 840 $\rm GeV$ in the combined $ej$ and $\mu j$ channels. 

On the other hand, a recent work from ATLAS \cite{atlastj} also presents a search for an $X$ new particle produced in association with a $t/\bar t$ quark, leading to the resonance in question. They found the data to be consistent with the SM expectation and excluded a particle with mass below 350 $ \rm GeV$ at 95 \% C.L, assuming unit right-handed coupling and null left-handed one. They also set limits in the mass-coupling plane at 95 \% C.L. for the case of a cross section scaling only as $g_{R}^{2}$. We show these limits in Fig.$~$\ref{all1}.

To conclude, CDF and CMS $tj/\bar t j$ resonance searches results do not affect the allowed area in the parameter space of the $Z'$ model defined by the triangle in Fig.$~$\ref{all1} while the analysis from ATLAS only excludes a small part of it leaving a considerable region of the parameter space safe from these searches.

Note that the results from $tj/\bar t j$ resonance searches can not be directly translated into constraints for a $Z'$ model with invisible decays since the increment of the width implies a cross section that no longer depends on the squared coupling only.

\subsection{Indirect constraints to the model}

Concerning the possible issues of the model, we have already mentioned its indirect constraints such as APV, $t \bar t j$ and single top production. Let us analyse each of them in the following paragraphs.

As it is well known, the model presented in this article may come into conflict with APV observables. In Ref.$~\cite{1203.1320}$ the limits given by APV have been studied in a model with a vector mediator coupled to $u_{R}$ and $t_{R}$ and a flavor-conserving boson. When we adapt their constraints to our model, we find the region compatible with APV limits for our model. These limits are given by the thick line in Fig.$~\ref{all1}$ for a cut-off $\Lambda=1000 ~ \rm GeV$. There are two main features of our model which, when contrasted to Ref.$~$\cite{1203.1320}, relax the APV constraints. The first one is that in our model the $Z'$ mass is larger than the top mass, and the second one is the fact that there is not a light flavor-conserving boson in our model. In any case, it is worth to mention at this point that the corrections to the calculation of the parity non-conservation in cesium are currently under discussion \cite{Dzuba:2012kx}.



We have also already referred to the {\it s-}channel processes that contribute to the $t \bar t j$ production as a difficulty of the model. To overcome it, we have proposed an increment in the $Z'$ width arguing that the new particle has invisible decays such as dark matter or neutrinos. As there are not available works on $t\bar tj$ limits that could be adapted to our model, we have used $W't d$ production results from Ref. \cite{1203.4489} as a rough estimation of the $t\bar tj$ production at the LHC in our model. We have checked that the $t \bar t j$ constraints in this work with $0.7 fb^{-1}$ are surpassed if the width is increased by a factor of 3. We have also noted that the $5 fb^{-1}$ {\it projected} constraints would not exclude the model if the increment of the width is of a factor of 5. Since $d$ quarks PDFs are different from those of the $u$ quarks, the analysis made in \cite{1203.4489} cannot be adapted to our model straightforward and needs a new study $\cite{progress}$.

The increment of the $Z'$ width brings with it an excess in single-top production. Single-top quarks can be produced through three different processes in the SM: a {\it t-}channel of the form $q b \to q t$ via the exchange of a $W$-like boson \cite{1103.2792},  a $W t$ associated production \cite{1005.4451} and an {\it s-}channel process \cite{1001.5034}. The {\it t-}channel process is dominant at both the LHC and the Tevatron. In our model, the single top production topology is given by one reconstructed top and missing energy, with no extra $b$ quark; different from that of the three processes mentioned. However, although the final state of the {\it t-}channel process and the $Wt$ associated production at L.O. do not have missing energy, they are the only processes of the three that do not have an extra $b$ quark. Henceforth, although the search strategy is not the same as that for the signature of our model, we use these processes as a reference to know how unlikely could be the excess in single top production predicted in our model. We use then, as an estimated reference, the latest measurements of the {\it t-}channel and $Wt$ associated single-top production. The ATLAS Collaboration results for both processes cross sections summed yield $\sigma_{t}=99.8 \pm 20.8$ pb \cite{1205.3130,1205.5764}, while those by CMS give $\sigma_{t}=92 \pm 15 $ pb \cite{cmssingletop,1201.4997}. In our model, for a $Z'$ width increased by a factor of 3, the expected excess in the single-top production cross section is $\sim (10-30)$ pb. In the case of the $Z'$ width increased by a factor of 5 (7) this excess reaches $\sim (10-40)$ ($\sim (15-40)$) pb. 

The sum of the {\it t-}channel and $Wt$ associated single top production cross section measured not only are in agreement with the predicted next-to-next-to leading order sum of $t$ and $\bar t$ cross sections for these processes, which is $\sigma_{t}=80.2 \pm 2.07$ pb \cite{1103.2792,1005.4451}, but also with our predictions for an increased width in most of the parameter space at 95 $\% $ C.L. The best agreement takes place when the $Z'$ width is increased by a factor of 3.

In order to perform a precise study of the single top production within our model with $Z'$ decays to invisible, a new search strategy should be developed. This new search strategy should involve one reconstructed top, missing energy and no extra $b$ quark \cite{progress}. 

At last, we should mention that, since the $t\bar t$ object is not created through the $Z'$ in a resonant channel in any of the NP diagrams, the model should not be constrained by the usual $t\bar t$ resonance searches.  Moreover, this model predicts a shape modification of the tail ($M_{t\bar t}\gtrsim 1-1.5$ TeV) of the spectrum, which is expected to be harder to measure than a resonant effect \cite{grojean}.

\subsection{Distinctive features of the model}

The most important distinctive feature of the model, and the cornerstone of this work, is the prediction of a large $A_{FB}$ and a small or even negative $A_{C}$ consistent with the experimental results from the Tevatron and the LHC in a region of the parameter space of the model. 

Let us now investigate some other specific features of the model which could be exploited in order to test it. To do so, it is interesting to study cuts on the charge asymmetry.

A first cut is in the transverse momentum of the $t\bar t$ pair, $p_{T}(t \bar t)$. As it is well known, in the SM one expects $A_{C}$ to grow with low $p_{T}$ and vice versa \cite{1202.6622}. In this model there is a particularity in the dependence of $A_{C}$ with $p_{T}(t \bar t)$ caused by the different {\it s-} and {\it t-}channel contributions to $A_{C}$ that results in different contributions to the $p_{T}(t \bar t)$. The {\it s-}channel processes involve a jet in the final state that provides the $t \bar t$ pair with an extra source of $p_{T}(t \bar t)$ apart from that of the initial state radiation (ISR). This is not the case for the {\it t-}channel process where the final state is $t \bar t$ and no additional sources of $p_{T}$ other than ISR exists. As a result, one expects the {\it s-}channel to be dominant for large values of $p_{T}(t \bar t)$. Since the {\it s-}channel contributes negatively to $A_{C}$, the model then predicts an excessive negative contribution to this observable when events with large $p_{T}(t \bar t)$ are considered. On the contrary, for low values of $p_{T}(t \bar t)$ the {\it t-}channel is preferred and one expects a smaller negative contribution from $A_{C_{s}}$ to the charge asymmetry, i.e., an excess in the positive contributions to $A_{C}$ coming from the {\it t-}channel processes. 
The first measurement of the dependence of $A_{C}$ with $p_{T}(t \bar t)$ for three bins in $p_{T}(t \bar t)$ is presented in Ref. \cite{cmsacy}.

An observation concerning the simulation of events with ISR is that the ISR modeling in Monte Carlo simulations has large uncertainties. These uncertainties are larger for $A_{C}$ in the low  $p_{T}(t \bar t)$ region. On the contrary, for large enough values of this variable the charge asymmetry becomes more independent of the ISR modeling, because the events passing such cuts in  $p_{T}(t \bar t)$ are dominated by gluon fusion events, which do not generate charge asymmetry. Henceforth, the prediction of a negative excess in $A_{C}$ for large values of $p_{T}(t \bar t)$ is not affected by large ISR modeling uncertainties. This is a prediction of our model. 

There is also another interesting cut on the charge asymmetry that could improve a NP search strategy. In this model the key channel is the {\it s-}channel, which involves a $qg$ collision and a $t\bar t j$ final state. In this kind of processes,  due to the presence of the jet, the $t \bar t$ pair is likely to have an extra source of $p_{T}$ apart from the already important contribution from initial state radiation of the incoming gluon. On the other hand, the incoming quark is likely to have considerably more momentum than the gluon so that the $qg$ events are more likely to be boosted in the $z$-direction, which translates into the $t \bar t$ pair having a large $p_{z}$ as well. As a result, one could improve the search strategy, in virtue of the proton PDFs, by requesting the $t \bar t$ pair not only to have large $p_{T}$ but also large $p_{z}$ simultaneously.

There are other variables that can contribute to the discovery of NP models at the LHC similar to the model studied in this work. For instance, Ref. \cite{1111.5857} investigates models where new $X$ mediators generate a charge-asymmetric signal in $t X$ production leading to observable new charge asymmetric variables in $t \bar t j$ events. Among these are $A_{C}$ as a function of the invariant mass and the transverse mass of various final state objects.



\section{Conclusions}

We have studied a phenomenological model with a new colorless, flavour-violating, electrically neutral $Z'$ boson with right handed couplings to $u$ and $t$ quarks. We assume that the $Z'$ is charged so that it is not the same as its conjugate partner. We also consider that this particle mass is larger than the top mass.

The interaction term 
$\bar utZ'$ 
and its hermitian conjugate with $Z'\neq Z'^{\dagger}$, introduce three new processes at L.O.: one in which the new particle is exchanged through a {\it t-}channel and the other two where it goes via an {\it s-}channel. We have observed that the {\it t-}channel process contributes positively both to $A_{FB}$ and $A_{C}$ and it is privileged at the Tevatron while the dominant {\it s-}channel one has a negative contribution and is turned on and only noticeable at the LHC, causing a cancellation of $A_{C}$ measured by this accelerator. This cancellation is not present at the Tevatron where actually a large $A_{FB}$ has been measured. We have studied the dynamics of the two processes involved in the cancellation in order to understand the mechanism through which it arises. This model then predicts a large positive $A_{FB}$ and a null or even negative $A_{C}$. 

We have studied the Tevatron and LHC phenomenology of this model and searched numerically for the cancellation mentioned. We investigated the allowed parameter space of the model by confronting it with several relevant and most stringent unfolded results from CDF and CMS at $95~\%$ C.L and with its major constraints such as $tj/\bar t j$ resonance searches, atomic parity violation and $t\bar tj$ and single top production.

We found that the direct constraint coming from CDF and CMS $tj/\bar t j$ resonance searches do not affect the allowed area in the parameter space of the model and that the same analysis from ATLAS only excludes a small part of it leaving a considerable region of the parameter space safe from these searches. On the other hand, we have also checked that the limits given by atomic parity violation are not in conflict with our model. Finally, since we found that the model predicts an excess in $t \bar t j$  final state, we have explored the possibility of increasing the $Z'$ width, assuming that the $Z'$ decays to not detectable particles, such as dark matter or neutrinos, a fraction of the times. With no available works on $t\bar tj$ limits that could be adapted to our model, we used $W't d$ production results from Ref. \cite{1203.4489} as a rough estimation of the $t\bar tj$ production at the LHC in our model. We have checked that the estimated $t \bar t j$ constraints are surpassed if the width is increased. We show that this increment of the $Z'$ width predicts single top production with a particular topology not present in the SM single top production: one reconstructed top and missing energy, with no extra $b$ quark. As a result, in order to know how unlikely could be the excess in single top production predicted in our model, we have used as a reference the two processes through which single tops are produced at L.O. in the SM that do not have a $b$ quark in the final state. We found the excess to be consistent with our predictions for an increased width in most of the parameter space at 95 $\% $ C.L. The best agreement takes place when the $Z'$ width is increased by a factor of 3. Let us mention that $t\bar tu$ constraints and single top search strategies studies are in progress. We expect that constraints coming from new results on $t\bar tj$ cross section will be those that place the tightest limits on our model.

At last, we have presented some distinctive features of the model by studying cuts to the charge asymmetry. We have noted that in this model the dependence of $A_{C}$ with $p_{T}(t \bar t)$ is caused by the different {\it s-} and {\it t-}channel contributions to $A_{C}$ which give rise to different contributions to the $p_{T}(t \bar t)$. This results in a prediction of an excessive negative contribution to $A_{C}$ when events with large $p_{T}(t \bar t)$ are considered and, on the contrary, an excess in the positive contributions to $A_{C}$ for low values of $p_{T}(t \bar t)$. We expect a study in the large $p_{T}(t \bar t)$ region to be more Monte Carlo independent than a study in the low $p_{T}(t \bar t)$ region.
We finally found that, because of the PDFs, one could improve the search strategy by requesting the $t \bar t$ pair not only to have large $p_{T}$ but also, simultaneously, large $p_{z}$.

 We have shown that our model brings compatibility to the apparent disagreement between $t\bar t$ Tevatron forward-backward asymmetry and LHC charge asymmetry. Let us finally remark that all the analysis has been made confronting our model with the Tevatron and LCH experimental results that seem to be more incompatible. If results with smaller apparent discrepancy were to be used, the constraints to the $Z'$ model would be less restrictive.

\subsection*{Acknowledgments}
We would like to thank J.~Adelman and M.~Peskin for useful discussions.  E.A.~thanks SLAC, where part of this work was carried out.

\end{document}